\documentstyle[emulateapj,apjfonts]{article}
\begin{document}

\title{Correcting for the Effects of Interstellar Extinction}
\author{Edward L. Fitzpatrick}
\affil{Department of Astronomy and Astrophysics, Villanova University,  Mendel Hall, Villanova, PA 19085}
\begin{abstract}

This paper addresses the issue of how best to correct astronomical data
for the wavelength-dependent effects of Galactic interstellar
extinction.  The main general features of extinction from the IR
through the UV are reviewed, along with the nature of observed spatial
variations.  The enormous range of extinction properties found in the
Galaxy, particularly in the UV spectral region, is illustrated.
Fortunately, there are some tight constraints on the wavelength
dependence of extinction and some general correlations between
extinction curve shape and interstellar environment.  These
relationships provide some guidance for correcting data for the effects
of extinction.   Several strategies for dereddening are discussed along
with estimates of the uncertainties inherent in each method.   In the
Appendix, a new derivation of the wavelength dependence of an average
Galactic extinction curve from the IR through the UV is presented,
along with a new estimate of how this extinction law varies with the
parameter $R \equiv A(V)/E(B-V)$.  These curves represent the
true monochromatic wavelength dependence of extinction and, as such, are
suitable for dereddening IR--UV spectrophotometric data of any
resolution, and can be used to derive extinction relations for any
photometry system.
 
\end{abstract}

\keywords{ISM:dust, extinction}

\section{Introduction}
 
A precise knowledge of the wavelength dependence of interstellar
extinction (i.e., the absorption and scattering of light by
interstellar dust grains) and of any spatial variability in this
dependence is important for two distinct reasons.  First, the
extinction depends on the optical properties of the dust grains along a
line-of-sight and potentially can reveal information about the
composition and size distribution of the grains.  Further, changes in
the extinction from place to place may reveal the degree and nature of
dust grain processing occurring in the ISM.  Second, the wavelength
dependence of extinction is required to remove the effects of dust
obscuration from observed energy distributions, since most astronomical
objects are viewed through at least some small amount of interstellar
dust.  Spatial variations in the extinction potentially limit the
accuracy to which energy distributions can be ``dereddened.'' Such
uncertainties might be acceptably small for very lightly reddened
objects, but quickly can become debilitating along modestly reddened
sightlines.

This paper follows from a talk given at the meeting ``Ultraviolet
Astrophysics Beyond the IUE Final Archive'' (Fitzpatrick 1998) and
addresses the second point raised above, i.e., the correction of energy
distributions for the effects of interstellar extinction.  The goals
are to provide a summary of what is known about the wavelength
dependence and spatial variability of interstellar extinction and to
present strategies for the best removal of the effects of extinction
from astronomical data.  In \S 2, I briefly review the chief features
of IR-through-UV extinction and the nature of the known spatial
variations.  These variations are illustrated with a set of optical/UV
extinction curves which span the (currently) known extreme limits of
extinction variations.  Constraints on the wavelength dependence of
extinction and some general correlations between extinction curve shape
and interstellar environment are also noted in \S 2.  Strategies for
dereddening are discussed in \S 3 along with estimates of the
appropriate uncertainties which ought to be incorporated into an error
analysis.  Some final comments regarding extinction along extragalactic
sightlines and at wavelengths shortward of 1150 \AA\/ are given in \S
4.  

The Appendix of this paper describes the construction of a new estimate
of the shape of a mean and an environment-dependent IR-through-UV
extinction curve.  These curves, which may be obtained from the author,
are suitable for dereddening multiwavelength spectrophotometric data,
such as has become available with the {\it Hubble Telescope's} Faint
Object Spectrograph, and for deriving extinction relationships for any
photometric system.

\section{Spatial Variations in Galactic Extinction}
\subsection{Variations Abound ...}
 
Figure 1 shows several estimates of the shape of a ``mean'' Galactic
extinction curve from the far-IR through the UV region.  The curves are
presented in a commonly used normalization scheme,
$E(\lambda-V)/E(B-V)$, and plotted against inverse wavelength.  The
significance of the different curves will be discussed later in this
paper.  For the present, these curves serve to illustrate the overall
characteristics of interstellar dust extinction.  Typically, the
extinction rises through the IR with a power law-like  dependence (see
Appendix), rolls over slightly in the optical region (``knee''), shows
a prominent feature at 2175 \AA\/ in the near-UV (``bump''), and has a
sometimes steep rise in the far-UV (``fuv rise'').  Figure 1 does not
attempt to show the relatively narrow optical/IR features which may be
associated with interstellar dust, such as the diffuse interstellar
bands in the optical or the solid state absorption (or emission)
features in the IR (e.g., the 9.8 \micron\/ silicate feature).

IR and optical extinction traditionally have been studied using ground
based broad- or intermediate- band photometry and
sightline-to-sightline variations have been recognized for a long time
(e.g., Johnson 1965).  These are often characterized using the
parameter $R \equiv A(V)/E(B-V)$, i.e., the ratio of total to selective
extinction at $V$.  The value of $R$ ranges between about 2.2 and 5.8
for sightlines along which UV extinction has also been measured, with a
mean value of $\sim$3.1 for the diffuse interstellar medium.  It is
believed that the shape of the IR extinction law longward of $\sim$8000
\AA\/ (the ``power law'' region) may be invariant and that the observed
range in $R$ is due to spatial variations in the steepness of the
extinction in the optical region around the ``knee''  (e.g., Cardelli,
Clayton, \& Mathis 1989;  Martin \& Whittet 1990).

Measurements of UV interstellar extinction have relied on satellite or
rocket spectrophotometric data, and date back to the original discovery
of the 2175 \AA\/ bump nearly 30 years ago (Stecher 1969; Bless \&
Savage 1970).  These studies immediately revealed that
sightline-to-sightline differences exist within the Galaxy in the
detailed properties of UV extinction curves.  Important results on the
degree of spatial variability of UV extinction were subsequently
derived from the large photometric databases accumulated by the {\it
TD-1} and {\it ANS} satellites (e.g.\, Koornneef 1978; Kester 1981;
Meyer \& Savage 1981).

The most graphic illustrations of spatial variations in UV extinction
were provided by the {\it IUE} satellite.  With its complete spectral
coverage between 1150 and 3200 \AA\/ and relatively high spectral
resolution  ($\sim$6 \AA\/ in the low-resolution mode) {\it IUE} was
well-suited for the study of UV extinction and many studies have
utilized {\it IUE} data to determine the wavelength dependence of
extinction along one or more Milky Way lines-of-sight (e.g.\, Bohlin \&
Savage 1980; Witt, Bohlin, \& Stecher 1980; Seab, Snow, \& Joseph 1981;
Morgan, McLachlan, \& Nandy 1982; Hecht et al. 1982; Massa, Savage, \&
Fitzpatrick 1983; Witt, Bohlin, \& Stecher 1984; Massa \& Savage 1984;
Franco, Magazz\`{u}, \& Stalio 1985; Aiello et al. 1988; and many
others).   Figure 2 shows analytical representations for $\sim$80 UV
extinction curves derived from {\it IUE\/} data (solid curves),
illustrating the wide range of properties observed for Galactic UV
extinction.  These curves are taken from the catalog of Fitzpatrick \&
Massa (1990), with a few additions, including the lines-of-sight toward
HD 210121 (Welty \& Fowler 1992) and HD 62542 (Cardelli \& Savage
1988).  The basis for the analytical representation will be discussed
further below.

The overall impression that has come from the last 30 years of
extinction studies is that there is a bewildering variety of
IR-through-UV extinction curves and that what has often been called
``peculiar'' extinction is actually quite common.  This bodes poorly
for the prospects of precisely dereddening astronomical energy
distributions.  In particular, the range of extinction properties seen
in the UV (Figure 2), where the effects of extinction are strongest,
may represent a bonanza of information for those who study the
properties of interstellar dust grains, but is a potential disaster for
those whose main interest in UV extinction is the removal of its
effects.

\subsection{... But There is Some Order Amid the Chaos}

The recognition that interstellar extinction is spatially highly
variable is critically important for both the study of dust grains and
for the purposes of dereddening.  Fortunately for the latter, it has
been realized that within the seemingly endless variety of observed
extinction curves there are actually some constraints on the wavelength
dependence and some links between extinction in the UV, optical, and
IR.

This began with Savage (1975) who showed, using {\it OAO-2} satellite
data, that the shape of 2175 \AA\/ bump can be reproduced well with a
Lorentzian profile.  This result --- which provides important clues
about the nature of the 2175 \AA\/ feature --- allows a functional
representation of the bump, and permitted Seaton (1979) to present an
estimate of the average Galactic UV extinction curve as a simple
analytical formula.  This curve is shown in Figure 1 by the dashed
curve, extending from 2.7 to 8.7 \micron$^{-1}$.

Following the lead of Savage and Seaton,  Fitzpatrick \& Massa (1986,
1988, 1990; collectively referred to hereafter as FM) examined {\it
IUE\/} extinction curves for many lines-of-sight and found that all
these curves could be fitted extremely well by a single analytical
expression with a small number of free parameters.  This expression
consists of (1) a Lorentzian-like bump term (requiring three
parameters, corresponding to bump width $\gamma$, position $x0$, and
strength $c3$), (2) a far-UV curvature term (one parameter $c4$), and
(3) a linear term underlying the bump and the far-UV (two parameters
$c1$ and $c2$).  This set of basis functions is shown in Figure 3.
With the proper choice of the 6 parameters, essentially all UV
extinction curves can be reproduced to within the uncertainties
inherent in the data.  For most sightlines, this number of free
parameters can be reduced to 4, since it has been seen that the
position of the 2175 \AA\/ bump is nearly invariant and that the
intercept and slope of the linear term are tightly correlated (FM; see
also Carnochan 1986 and Jenniskens \& Greenberg 1993), such that the
linear term appears to pivot about the point
[$1/\lambda,\frac{E(\lambda-V)}{E(B-V)}$] $\simeq$ [3,2].  The UV
portions of the curves shown in Figure 2 ($\lambda < 2700$ \AA) were
all constructed using these functions.  There is, thus far, no convincing
evidence for substructure in UV extinction curves.
 
The FM results show that some order underlies the wide spectrum of
wavelength dependences seen in the UV, but do not necessarily permit a
more accurate dereddening of an object along any given line of sight.
Fortunately it has been observed that extinction curves do not vary
randomly across the sky, but show distinct regional signatures (e.g.,
Meyer \& Savage 1981; Morgan, McLachlan, \& Nandy 1982; Panek 1983;
Massa \& Savage 1984; Clayton \& Fitzpatrick 1987) and it is clear that
extinction properties reflect the physical conditions and past
processing histories of the environments in which the dust grains
reside.

The work of Cardelli, Clayton, \& Mathis (1988, 1989; hereafter CCM)
provides a link between one measure of dust grain environment and the
wavelength dependence of UV-through-IR extinction.  CCM found that the
shape of UV extinction curves correlates with the parameter $R$
(defined in \S 2.1 above).  This suggests that --- although there is
considerable scatter --- extinction curves from the UV through the IR
can be characterized as approximately a one-parameter family dependent
on $R$.  Thus, if the value of $R$ can be determined (from optical and
IR photometry), then the properties of the the entire UV--IR extinction
curve can be predicted.  The essence of the CCM result is illustrated
in Figure 4.  Four representative extinction curves are shown, each
determined by the value of $R$ listed at the righthand side of the
curves.   This figure shows that extinction curves which are seen to be
very ``flat,'' or ``grey,'' in the UV roll over strongly in the optical
and are characterized by large values of $R$.  Steep UV curves remain
steep in the optical and are characterized by small values of $R$.  The
CCM curve for $R=3.1$ is shown in Figure 1 by the dotted line.

The results of CCM are important in 3 ways:  (1) They indicate that
some of the spatial variations seen in extinction curves behave
coherently and systematically over a wide wavelength range, thus
potentially allowing for the consistent dereddening of multiwavelength
energy distributions.  (2) The observed dependence on $R$ allows for
the definition of a meaningful average Galactic extinction curve.
Existing datasets of extinction curves, particularly for the UV, are
biased toward ``interesting'' regions where extinction properties are
extreme.  Computing a simple mean curve from these data does not
necessarily yield a reasonable estimate of a Galactic mean.  Since it
is well-established, however,  that the appropriate mean value of $R$
for the diffuse ISM is $\sim$3.1 (e.g., Schultz \& Wiemer 1975; Whittet
\& van Breda 1980; Rieke \& Lebofsky 1985), a reasonable definition of
a mean Galactic extinction law is that which corresponds to the case
where $R = 3.1$.  (3)  They demonstrate a general correlation between
dust grain environment and the wavelength dependence of extinction,
since large values of $R$ are generally found in dense environments
where dust grain growth is thought to occur.  The ``greyness'' of the
UV/optical extinction curves for large $R$ is consistent with a larger
than normal dust grain population.   These results have an important
bearing on how to correct for the effects of extinction, discussed in
the following section.

\section{Correcting for the Effects of Extinction}

Given the observed complexity of Galactic UV extinction, what is the
best way to deredden an observed energy distribution?   There are three
different possibilities, which involve the use of a global mean
extinction curve, an $R$-dependent curve, and a sightline-specific
curve.  These three alternatives and their inherent uncertainties are
discussed below, and a new derivation of the shape of the average
IR-through-UV extinction curve is presented.  A more thorough
discussion of the uncertainties involved in UV extinction corrections
is given by Massa (1987).

\subsection{Using a Global Mean Curve}

If there is no specific information available about the wavelength
dependence of the extinction along a sightline of interest then the
only alternative is to adopt some globally defined mean curve and
perform a realistic error analysis.  This is the least attractive of
the three cases discussed here, but is by far the most commonly used.
The average Galactic extinction curves from either Seaton (1979; see
Figure 1) or Savage \& Mathis (1979) are often adopted for dereddening
UV data.  These should now be superseded by the use of an $R$-dependent
curve computed for the case $R = 3.1$ (see \S 2.2).  Figure 1 shows the
$R = 3.1$ curve from CCM (dotted line) and a new determination of the
$R = 3.1$ case (solid line) which is described in the Appendix to this
paper.  This new curve aims to reproduce the detailed wavelength
dependence of the $R = 3.1$ extinction law and has been constructed to
account for bandpass effects properly in optical/IR extinction data and
to reproduce the observed broad-, intermediate-, and narrow- band
extinction measurements.  The curve is thus suitable for dereddening
multiwavelength spectrophotometric observations and can be used to
derive the average extinction relationships in any photometric system.

A complete evaluation of the likely error in a dereddened relative
energy distribution $m(\lambda-V)$ requires the propagation of the
uncertainty in $E(B-V)$ and the often-neglected but often-dominant
uncertainty in the adopted mean curve.  This is given by
\begin{equation}
\sigma^2_{m(\lambda-V)} =  \sigma^2_{E(B-V)} \times k(\lambda-V)^2 + E(B-V)^2 \times \sigma^2_{k(\lambda-V)}
\end{equation}
where $k(\lambda-V)$ represents the normalized extinction curve
$E(\lambda-V)/E(B-V)$.  If absolute fluxes are desired, then the
uncertainty in the assumed value of $R$ must also be incorporated (two
additional terms on the righthand side of eq. 1, similar to the current
terms but with $R$ substituted for $k(\lambda-V)$).

The large database of {\it ANS} satellite extinction measurements from
Savage et al. (1985) shows that the 1-$\sigma$ scatter at 1500 \AA\/ is
$\sigma_{k(15-V)} = 0.74$, based on $\sim$400 sightlines with $E(B-V)
\ge 0.5$.  (Only relatively large values of $E(B-V)$ were considered in
order to minimize the effects of spectral mismatch error and random
noise on the $\sigma$ measurement.) To extend this analysis to other
wavelengths, we computed the standard deviation at each wavelength
point for the 80 curves shown in Figure 2, and then scaled the result
to match the {\it ANS} value at 1500 \AA.  (The actual standard
deviation at 1500 \AA\/ for the 80 {\it IUE} curves is somewhat higher
than the {\it ANS} result because of the bias in Figure 2 toward
extreme extinction curves).  The resultant values of
$\sigma_{k(\lambda-V)}$ are shown by the thick dotted curve near the
bottom of Figure 2 (labeled ``$\sigma$'') and are listed at selected
wavelengths in the third column of Table 1.   This estimate of
$\sigma_{k(\lambda-V)}$ should be adopted whenever the average Galactic
extinction curve is used for dereddening an observed energy
distribution.  The uncertainties approach zero for $1/\lambda < 3$ $\mu
m^{-1}$ due to the curve normalization.  The quantity $E(B-V)$ is
usually derived directly from photometry and often has an easily
quantifiable uncertainty; thus the computation of
$\sigma^2_{m(\lambda-V)}$ from eq. 1 is straightforward.

Sometimes $E(B-V)$ is not known {\it a priori}\/ and is estimated  by
``ironing out'' the 2175 \AA\/ extinction bump using an assumed
extinction curve shape (e.g., see Massa, Savage, \& Fitzpatrick 1983).
In many cases, uncertainties in $E(B-V)$ as small as 0.01-0.02 mag for
moderately reddened objects have been quoted from this process.  This
is incorrect!  The normalized ``height'' of the 2175 \AA\/ extinction
bump --- which is the quantity that is important in the ironing-out
process --- has a 1-$\sigma$ scatter of about $\pm$20\% around its mean
value (from the data in Figure 2 and Savage et al. 1985).  Thus the
uncertainty in an $E(B-V)$ measurement derived from the bump must be
considered to have a similar relative uncertainty.  Since bump strength
does not correlate well with other aspects of UV extinction, such as
the slope of the linear component or the strength of the far-UV rise
(FM), the uncertainty in an energy distribution dereddened this way can
be estimated by using eq.  1 with $\sigma_{E(B-V)} \simeq 20\%$ and the
values of $\sigma_{k(\lambda-V)}$ from Table 1.  The result of this
calculation needs to be modified somewhat by removing the signature of
the 2175 \AA\/ bump.  The final estimate of $\sigma^2_{m(\lambda-V)}$
computed this way is listed in the last column of Table 1.  Multiplying
these values by $E(B-V)$ gives the uncertainty in a dereddened relative
energy distribution.  The values in Table 1 agree well with estimates
$\sigma^2_{m(\lambda-V)}$ at 1500, 1800, 2200, 2500, and 3300 \AA\/
derived from the {\it ANS} data of Savage et al (1985), and are
consistent with the analysis of Massa (1987).

\subsection{Using a $R$-Dependent Mean Curve}

If the value of $R$ has been measured along a sightline of interest, or
if some general information is available about the dust grain
environment (e.g., dense region or diffuse region?), then the
uncertainties in dereddening can be reduced somewhat by adopting an
$R$-dependent extinction curve computed at the appropriate value of
$R$.  For the original 29 sightlines used by CCM, the standard
deviation of the individual observed extinction curves minus the
computed $R$-dependent curves is about a factor of 2 smaller than when
the individual curves are compared to the average Galactic ($R = 3.1$)
curve.  Thus the uncertainties in a dereddened energy distribution can
be estimated using eq. 1 and adopting 0.5 times the
$\sigma_{k(\lambda-V)}$ values from Table 1.  If $R$ is poorly measured
or only estimated from environmental factors, then allowance must be
made for this in the error analysis.

Using an appropriate $R$-dependent curve can slightly improve the
accuracy of results from ironing out the 2175 \AA\/ bump.  The values
of $\sigma^2_{m(\lambda-V)}$ can be estimated as above in \S 3.1, using
the same value of $\sigma_{E(B-V)} \simeq 20\%$ and 0.5 $\times$
$\sigma_{k(\lambda-V)}$ from Table 1.  The resultant values of
$\sigma^2_{m(\lambda-V)}$ are about 85\% as large as those listed in
the last column of Table 1.  The gain from using the $R$-dependent
curves is only marginal because the normalized bump heights are highly
variable and not correlated with $R$.

It is important to note that, even when $R$ and $E(B-V)$ are
well-determined, the use of an $R$-dependent extinction curve only
reduces --- but does not eliminate --- the wavelength dependent
dereddening error.   The standard deviation of the observed vs.
$R$-dependent curves does not go to zero, even for the sample of 29
sightlines used by CCM to define the $R$-dependence, because the
UV/optical curves are not really a one-parameter family dependent only on
$R$.   Curves derived for different sightlines with the same values of
$R$ show a wide range of properties, including differences in the
strengths of the bump and the far-UV rise.  Thus, as noted by CCM, the
CCM formula reproduces a general trend, but does not provide
particularly good fits to individual extinction curves, even when the
value of $R$ is well-determined.  

Apart from this intrinsic scatter around the mean $R$ relation, the
accuracy of the CCM results is limited in the IR/optical region due to
bandpass effects with the broadband Johnson filters used to measure the
extinction.   As a result, CCM tends to overestimate the level of
extinction in the near-IR and blue-visible.  In the Appendix, a new
derivation of the $R$-dependence of IR-through-UV extinction is
presented.  This corrects for the systematic bandpass effects in CCM,
but is still plagued by the same uncertainties caused by the ``cosmic''
scatter of extinction properties around the mean $R$ relation.

\subsection{Using a Sightline-Specific Curve}

Without question, the best of all possible dereddening choices is to
use an extinction curve derived explicitly for the sightline of
interest.  This advice is not so unhelpful as it may sound, because the
archives of the {\it IUE\/} satellite  may well contain normal stars in
the direction (and at the distance) of the object of interest, from
which extinction curves may be derived (e.g., halo stars in the case of
extragalactic sightlines).  The accuracy of the dereddened energy
distribution may thus be limited chiefly by the uncertainties in the
``pair method'' of constructing the extinction curve (see, e.g., Massa
et al. 1983) and by the determination of $E(B-V)$ for the object of
interest.  For a well-determined curve, ``ironing out'' the bump --- in
the cases when $E(B-V)$ is not known explicitly --- should yield the
same degree of accuracy as for the cases when $E(B-V)$ is known {\it a
priori}.  An example of such an approach is the determination of the
intrinsic energy distributions of rare or exotic objects located in
clusters with more mundane objects.  Massa \& Savage (1985) utilize
this technique, using extinction curves derived from main sequence B
stars in open clusters to determine the continuum shapes of cluster O
stars.

\section{Final Comments}

The results discussed above strictly pertain to Milky Way extinction
for wavelengths longward of about 1150 \AA\/ (the {\it IUE} satellite
cutoff).  Observers of extragalactic objects must contend with
foreground Milky Way extinction as well as extinction due to dust
grains local to the object of interest.  It is reasonable to assume
that the properties of interstellar extinction in external galaxies are
fully as complex as found for the Milky Way.  However, only for the
nearby Large Magellanic Cloud (LMC) do observed extinction curves for
individual sightlines approach the accuracy needed to study spatial
extinction variations (see Fitzpatrick 1998 and references within).
The properties of Milky Way extinction along the high galactic latitude
sightlines relevant to most extragalactic observations are also not
well-determined.  One of the few quantitative studies of high latitude
extinction, that by Kiszkurno-Koziej \& Lequeux (1987) utilizing {\it
ANS} satellite data, suggests that halo extinction is slightly steeper
in the far-UV and has a weaker 2175 \AA\/ bump than the typical
extinction found in the plane of the Galaxy.  Attempts to eliminate
galactic foreground reddening by ironing out the bump may lead to
systematic underestimates of the total amount of foreground
extinction.

The properties of extinction in the wavelength range 912-1150 \AA\/ are
poorly known, but the work that has been done (e.g., York et al. 1973;
Snow, Allen, \& Polidan 1990) suggests that the far-UV rise continues
at least down to wavelengths of 950 \AA\/ or so.  An intrinsic problem
with measuring extinction in this wavelength range is the high density
of H I and H$_2$ absorption lines which mimic continuous absorption in
low-resolution data.  CCM suggest that the wavelength dependence of
extinction in this region is reasonably well-represented by an
extrapolation of the fitting functions used in the {\it IUE} wavelength
range.  At the present, the best dereddening strategy for 912-1150
\AA\/ data would be to adopt such an extrapolation, based if possible
on {\it IUE} extinction curves derived for the sightline of interest.
The uncertainties listed in Table 1 for 1000 \AA\/ are based on this
simple extrapolation.

%
%

\acknowledgements
 
I thank my partner-in-interstellar-grime, Derck Massa, for many years
of enjoyable collaboration.  I also acknowledge the contributions of
the late Jason Cardelli to the study of interstellar extinction.
Future progress in the field will proceed at a slower pace without
Jason's enthusiasm and scientific insight.  This work was supported in
part by NASA grant no. NAG5-7113.

%
%

\begin{appendix}
\section{Deriving a $R$-Dependent IR-through-UV Extinction Curve}

In this Appendix a new estimate of the wavelength dependence of Milky
Way extinction from the IR through the UV is derived.  The motivation
for this is the need for a multiwavelength extinction curve suitable
for dereddening IR-through-UV spectrophotometric data now available
from, for example, the  {\it Hubble Space Telescope's} Faint Object
Spectrograph (see, e.g., Fitzpatrick \& Massa 1998 and Guinan et al.
1998).  The specific goals are 1) to produce a curve for the $R=3.1$
case which reproduces existing narrow band extinction measurements {\it
and} the principal Johnson and Str\"{o}mgren photometric extinction
relations and 2) to define the $R$-dependence of the extinction law,
relying principally on spectral regions where this dependence is
well-determined.   Existing estimates of the average wavelength
dependence of optical/near-IR extinction do not accurately portray the
true shape of the curve due to bandpass effects arising from the
broadband photometry used to derive the curves.  This deficiency will
be corrected by using synthetic photometry. 

\subsection{The UV Region}

CCM noted a correlation between the value of $R^{-1}$ (as determined
from IR/optical photometry) and the values of $A(\lambda)/A(V)$ (i.e.,
total extinction at $\lambda$ normalized by total extinction at V) at
UV wavelengths using parametrized extinction curves for a subset of 29
stars from the FM sample.  They consequently derived a complex
polynomial expression to reproduce the wavelength- and $R$-
dependences.  We approach the problem slightly differently and note
that the essence of the CCM result is two correlations involving the
coefficients of the FM fitting function originally used to parametrize
the extinction curves studied by CCM.  We thus derive the functional
forms of these correlations and retain the FM fitting function to
compute the values of $E(\lambda-V)/E(B-V)$ for wavelengths $\lambda <
2700$ \AA.
 
Figure 5 illustrates the two correlations.  The top panel shows a
plot of the FM parameter $c2$ (representing the slope of the linear UV
extinction component) vs. $R^{-1}$ (filled circles) along with a least
squares estimate of the linear relationship between the two (dashed
line).  The data are for a subset of the FM sample for which IR
photometry could be used to deduce the value of $R$ (see
below),  with the addition of the sightline toward HD 210121 (Welty \&
Fowler 1992; Larson, Whittet, \& Hough 1996).  This represents the only
convincing correlation between $R$ and the properties of the UV
extinction curve (see, e.g., Jenniskens \& Greenberg 1993).  The
equation of the least-squares fit to the data is
\begin{equation}
c2 = -0.824 + 4.717\times R^{-1}             \\
\end{equation}
No physical interpretation or significance is placed on the form of the
adopted functional relationship between $R$ and $c2$; it is merely that
which best reproduces the observed correlation.  We note, however, that
the intrinsic slope of extinction in the optical region is proportional
to $R^{-1}$, and that eq. A1 thus simply implies that the slopes of the
extinction curves in the UV and optical vary together in a linear
manner --- as optical extinction steepens, UV extinction steepens.
 
The bottom panel of Figure 5 shows the well-known relationship between
slopes ($c2$) and intercepts ($c1$) of the linear background component
(FM; Carnochan 1986; Jenniskens \& Greenberg 1993).  The data are for
the full set of 80 curves from FM, plus HD 210121, and minus the Orion
Nebula stars (HD 36982, 37022, 37023, and 37061) which suffer from
scattered light contamination (which mainly affects the linear intercept).
The equation of the least squares fit to this relationship (dashed
line), is
\begin{equation}
c1 =  2.030 - 3.007 \times c2                       \\
\end{equation}

We use the following mean values for the other four parameters required
to specify UV extinction curves with the FM formula:  $x0$ (bump
position) = 4.596 \micron$^{-1}$; $\gamma$ (bump width) = 0.99
\micron$^{-1}$; $c3$ (bump strength) = 3.23; and $c4$ (FUV curvature) =
0.41.

\subsection{The Optical/IR Region}

A common way of representing the wavelength dependence of extinction in
the optical and IR regions is to draw a smooth curve through the values
of $E(\lambda-V)/E(B-V)$ (or some other curve normalization) derived
from Johnson {\it UBVRIJHKLM} photometry, assuming that these represent
the monochromatic values of the extinction at the filter effective
wavelengths $\lambda_{eff}$ (e.g., CCM, Martin \& Whittet 1990).  This
procedure {\it does not} accurately yield the monochromatic extinction
curve because it ignores the wavelength dependence of extinction across
the width of an individual filter.  Extinction decreases towards longer
wavelengths in the optical/IR region and therefore the value of
$\lambda_{eff}$ for a reddened star is shifted to a longer wavelength
than for an identical unreddened star.  An extinction measurement made
by comparing photometric indices for two such stars (i.e., the ``pair
method'') will always overestimate the monochromatic extinction in the
neighborhood of $\lambda_{eff}$ because stellar emergent fluxes in the
optical/IR decrease towards longer wavelengths (for early-type stars).
The magnitude of this effect depends on how much the true extinction
curve varies across the filter.  A broadband $E(\lambda-V)/E(B-V)$
measurement will also be influenced by the intrinsic energy
distributions of the stars used in the pair method and by the total
amount of extinction.
 
The approach taken here is to find the wavelength dependence of the
optical/IR extinction curve which reproduces the photometric extinction
measurements when synthetic photometry of an artificially reddened
stellar energy distribution is compared with that of identical but
unreddened star.  To represent the stellar energy distribution, an
ATLAS9 model atmosphere from R.L.  Kurucz (with $T_{eff} = 30000$ K and
$\log g = 4.0$) is used, along with an adopted value of $E(B-V) =
0.5$.  Synthetic photometry is performed to yield measurements in the
broadband Johnson {\it UBVRIJHKLM} system and the intermediate
band Str\"{o}mgren {\it uvby} system.  The first step is to determine the
curve shape for the mean $R=3.1$ case and then to define the nature of
the variation with $R$.  The first two columns of Table 2 list the
photometric extinction ratios (in both the Johnson and Str\"{o}mgren
systems) and their observed values, which the $R=3.1$ extinction curve
is required to reproduce.  References to the observations are given in
Column 3.  In addition, the curve is constrained to reproduce the mean
narrowband measurements in the 3400--7900 \AA\/ region published by
Bastiaansen (1992), which are assumed to represent the monochromatic
values of the extinction for the case $R = 3.1$ at the filter central
wavelengths.

Figure 6 shows the shape of the monochromatic extinction curve which
best satisfies these requirements (thick solid curve), plotted as total
extinction $A(\lambda)$ normalized by $E(B-V)$.  The arbitrarily scaled
profiles of the Johnson and Str\"{o}mgren filters are indicated, and
the Bastiaansen data shown by the small plus signs (``+'').  The
agreement between the $R=3.1$ curve and the Bastiaansen data is clear
from the figure.  The fourth column of Table 4 gives the values of the
photometric extinction ratios produced by the curve, also in very good
agreement with the observations.

For wavelengths longward of 2700 \AA ($1/\lambda < 3.7 \mu$m$^{-1}$),
the $R=3.1$ curve is defined by a cubic spline interpolation between a
set of optical/IR anchor points (filled circles) and a pair of UV
anchor points (filled squares).  The values of the UV anchors (at 2700
\AA\/ and 2600 \AA)  are determined by the FM fitting function for the
case $R=3.1$ (see above) and assure a smooth junction between the
optical and UV regions at 2700 \AA.  The anchor at 0 \micron$^{-1}$ is
fixed at 0 (i.e., no extinction at infinite wavelength) and the values
of the other six optical/IR points were adjusted iteratively to find
the curve shape which best reproduced the extinction observations.  The
wavelengths chosen for the spline anchor points are somewhat arbitrary,
although points in the IR, the optical normalization region, and the
near-UV are clearly required.  The wavelengths and values of the spline
anchors for the $R=3.1$ curve are given in Table 3. 

Note the {\em slope} of the derived extinction curve does not approach
zero as $1/\lambda$ approaches zero.  We sacrificed this physically
expected requirement in order to achieve a better fit to the Johnson IR
photometry and to preserve the simplicity of the fitting procedure.  A
zero slope could have been guaranteed by, for example, adopting a power
law to represent the extinction; but no single power law can reproduce
the IR photometry to an acceptable level.  The new curve should be
treated as very approximate at wavelengths beyond the limit of the
M band (i.e., at $\lambda > 6 \micron$).

The overall $R$-dependence of the optical/IR curve is relatively easy
to incorporate by the following adjustments in the spline anchors:  (1)
the UV points are computed by the FM fitting function using the
coefficient values given above, including the $R$-dependent value of
$c2$;  (2) the IR points at $1/\lambda < 1 $\micron$^{-1}$ are simply
scaled by $R$/3.1, since the shape of the far-IR extinction is believed
to be invariant (see \S 2.1 and below); and (3) the optical points are
vertically offset by an amount $R-3.1$, with slight corrections made to
preserve the normalization.  Without these corrections (which are less
than 0.015 over the range $R$ = 2 to 6) the extinction curves would
drift away from the standard normalization, i.e., $E(B-V) = 1$, by a
few hundredths of a magnitude as $R$ departed from the value 3.1.
Table 4 gives formulae for computing the $R$-dependent values of the
optical spline anchors.  The corrections just noted are manifested
in the departure of the linear term from a value of 1.0 and in the
higher order term for the 4110 \AA\/ point.

It is not obvious {\em a priori} that the above adjustments in the
spline should produce the proper curve shapes in the relatively large
gaps between the 4 points in the optical normalization region and the
UV or IR regions. However, it will be shown below that the resultant
curves are in good agreement with observations.

\subsection{The Full {\it R}-Dependent Curve}

Figure 7 shows the full wavelength range of the IR-through-UV
$R$-dependent extinction curves derived here, for four representative
values of $R$ (thick solid and dashed lines).  For $\lambda \le 2700$
\AA\/ the curves are computed using the FM fitting function and for
$\lambda >  2700$ \AA\/ the curves are spline interpolations
between the $R$-dependent spline anchors.   The CCM results are shown
for comparison by the thin dotted lines for the same four $R$ values.
In the UV region, the new results and the CCM curves are similar for $R
\le$ $\sim$4.0, but diverge for larger $R$.  The main reason for this
discrepancy lies in the UV linear extinction component.  While CCM
adopted a linear relationship between the slope $c2$ and intercept
$c1$,  their relation --- shown by the dotted line in the bottom panel
of Figure 5 --- does not agree well with the observations.  The large
discrepancy at small values of $c2$ (i.e., large $R$) produces the
difference seen in Figure 7.
 
A closeup comparison between the new results and CCM in the optical/IR
region for the $R=3.1$ case can be seen in Figure 6, where the CCM
curve is indicated by the dotted line.  The disagreement in the region
near the Johnson $R$ filter illustrates the bandwidth effect discussed
above.  The CCM curve is fixed at $A(\lambda)/E(B-V) = 2.32$ at 7000
\AA, which is taken as $\lambda_{eff}$ for the Johnson $R$.  The new
curve is constructed to yield the same value of $A(\lambda)/E(B-V)$ but
for synthetic photometry with the Johnson $R$.  The discrepancy is
particularly large for this filter because of the steep change in
extinction across its relatively broad profile.  This effect also
accounts for the lesser discrepancies in the regions of the $U$, $B$,
$I$, and $J$ filters.  O'Donnell (1994) presented a revision to the CCM
formula in the region of the $U$, $B$, and $V$ filters based on the
Str\"{o}mgren photometric indices.  This revision uses the same stategy
as CCM of fixing the extinction ratios at the filter effective
wavelengths, but more closely resembles the newly derived monochromatic
extinction curve because the bandwidth effects in the narrower
Str\"{o}mgren filters are smaller.

Note that the technique used here to produce $R$-dependent curves can
be used to construct ``customized'' UV/optical extinction curves for
sightlines with well-defined UV extinction properties.  It is simply
necessary to substitute the measured values of $A(\lambda)/E(B-V)$ at
2700 \AA\/ and 2600 \AA\/ for the UV spline anchors listed in Table 3
and then perform the cubic spline interpolation to determine the
optical portion of the curve, which smoothly joins the measured UV
curve at 2700 \AA.  This method was used in Figure 2 to extend the
{\it IUE} curves into the optical region.

The assertion that the shape of the IR extinction law is invariant can
be tested by comparing results obtained from the $R$-dependent curves
in Figure 7 --- which were constructed based on this assumption ---
with observations.  This comparison is made in Table 5 for a number of
sightlines which span the observed range in $R$.  The ``model'' data
listed for each sightline are the results from synthetic photometry on
model stellar energy distributions artificially reddened using the
$R$-dependent curves, with $E(B-V)$ values from column 2 of the table
and $R$ values from column 7.  The listed values of $R$ are those that
best reproduce the observed color excesses and the uncertainties result
from assuming uncertainties of $\pm$0.02 mag in $E(B-V)$ and $\pm$0.05
mag in the IR $E(V-\lambda)$.  The agreement between observations and
model values is excellent and there are no strong systematic trends
evident.  The results are thus consistent with an invariant IR
extinction curve.  At wavelengths greater than $\sim$1 \micron, the
extinction curve roughly resembles a power law with an index of $\sim$1.5.
This is similar to that adopted by CCM (1.6), but much flatter than
that of Martin \& Whittet (1990; $\sim$1.8).

Values of $R$ are sometimes estimated from the relation $R \simeq 1.1
\times E(V-K)/E(B-V)$.  Exact relationships between R and the IR color
excesses can be derived for the new $R$-dependent curves.  These are
given by
\begin{equation}
R = 1.39 \times \frac{E(V-J)}{E(B-V)} - 0.02                       \\
\end{equation}
\begin{equation}
R = 1.19 \times \frac{E(V-H)}{E(B-V)} + 0.04                       \\
\end{equation}
\begin{equation}
R = 1.12 \times \frac{E(V-K)}{E(B-V)} + 0.02                       \\
\end{equation}
\begin{equation}
R = 1.07 \times \frac{E(V-L)}{E(B-V)} - 0.01                       \\
\end{equation}
The coefficients in each of these equations actually depend on
the value of E(B-V) itself but, over the range $E(B-V) = 0$ -- 2.0,
vary by only several hundredths.  The values in the equations are the
results for $E(B-V) = 0.5$.  
 
At the blue end of the optical region, the interpolation between the
optical and UV spline anchor points can be tested by comparing
predictions of the extinction indices $E(U-B)/E(B-V)$ and
$E(c1)/E(b-y)$ with photometric measurements.  Figure 8 shows this
comparison, with the difference between the observed values and
predicted values --- derived from synthetic photometry of
``customized'' extinction curves from the FM sample --- plotted against
the slope of the linear component $c2$.  The curves were produced as
described above, by adopting the observed values of $A(\lambda)/E(B-V)$
at 2600 \AA\/ and 2700 \AA\/ as the UV anchor points. The random scatter
(measurement noise) in both indices is large,  but no
systematic trends are seen and the model curves appear to reproduce at
least the integrated properties of extinction well in the near-UV
region. Over the range of $c2$ values shown, the model values of
$E(U-B)/E(B-V)$ range from about 0.6 to 0.8 and the values of
$E(c1)/E(b-y)$ from about 0.0 to 0.3.

In summary, the $R$-dependent curves derived here reproduce existing
photometric and spectrophotometric measurements and provide a good
estimate of  the true monochromatic wavelength dependence of
interstellar extinction in the IR-through-UV regions.  These curves
should be preferred for dereddening UV, optical, and near-IR
spectrophotometry.  Since the new curves give the detailed wavelength
dependence of extinction, they can be used to predict the extinction
relationships for any photometric system by using synthetic photometry
of artificially reddened energy distributions.  An IDL procedure to
produce the new curves at any desired value of $R$ and over any
wavelength range can be obtained from the author or via anonymous ftp
at {\it astro1.vill.edu}.  After logging in, change directories to {\it
pub/fitz/Extinction}, and download the file ``FMRCURVE.pro.''
Alternatively, this directory contains a series of compressed files
named ``FMRCURVEn.n.txt'' (e.g., ``FMRCURVE3.1.txt'') which contain
ASCII versions of the curves for various values of $R$ (``n.n'').  Two
columns of data are contained in each file; the first contains
wavelengths in \AA\/ and the second contains the extinction curve in
$A(\lambda)/E(B-V)$. There are 1099 sets of points in each file.

\end{appendix}

%
%


%
%


\footnotesize
\begin{deluxetable}{cccc}
\tablenum{1}
\tablewidth{0pc}
\tablecaption{Uncertainties in Extinction Corrections}
\tablehead{
\colhead{}                                                 & 
\colhead{}                                                 &
\colhead{$\sigma_{k(\lambda-V)}$\tablenotemark{a}}         & 
\colhead{$\sigma_{m(\lambda-V)}/E(B-V)$\tablenotemark{b}}  \\  
\colhead{Wavelength}                                       & 
\colhead{$\lambda^{-1}$  }                                 &
\colhead{(for corrections using}                           & 
\colhead{(for ``ironing out'' }                            \\ 
\colhead{(\AA)}                                            & 
\colhead{($\mu m^{-1})$}                                   & 
\colhead{average Galactic curve)}                          & 
\colhead{the 2175 \AA\/ bump)}                             \\
\colhead{(1)} &
\colhead{(2)} &
\colhead{(3)} &
\colhead{(4) }}
\startdata
1000 &     10.0 & 2.5\phn      & 3.8\phn    \nl
1100 &  \phn9.1 & 1.7\phn      & 2.8\phn    \nl
1200 &  \phn8.3 & 1.3\phn      & 2.1\phn    \nl
1300 &  \phn7.7 & 1.0\phn      & 1.7\phn    \nl
1400 &  \phn7.1 & 0.86         & 1.5\phn    \nl
1500 &  \phn6.7 & 0.74         & 1.3\phn    \nl
1600 &  \phn6.2 & 0.66         & 1.1\phn    \nl
1800 &  \phn5.6 & 0.55         & 1.0\phn    \nl
2000 &  \phn5.0 & 0.55 & 0.89    \nl
2200 &  \phn4.5 & 0.64 & 0.78    \nl
2400 &  \phn4.2 & 0.43 & 0.69    \nl
2600 &  \phn3.8 & 0.30 & 0.62    \nl
2800 &  \phn3.6 & 0.23 & 0.55    \nl
3000 &  \phn3.3 & 0.17 & 0.48    \nl
3500 &  \phn2.9 & 0.06 & 0.35    \nl
4000 &  \phn2.5 & 0.00 & 0.25    \nl                                                    \nl
\tablenotetext{a}{$k(\lambda-V) \equiv E(\lambda-V)/E(B-V)$.  These
values of $\sigma_{k(\lambda-V)}$ should be inserted into Eq. 1 to
compute the uncertainty in a corrected relative energy distribution
$m(\lambda-V)$ when there is no information available about the true
shape of the extinction law along the sightline and the average
Galactic curve is adopted by default.  See \S 3. in the text.}
\tablenotetext{b}{These values (multiplied by the derived $E(B-V)$) give
the uncertainty in the final relative energy distribution
$m(\lambda-V)$ when an object is dereddened by ``ironing out'' the 2175
\AA\/ bump using the average Galactic extinction curve. See \S 3.1 in
the text.}
\enddata
\end{deluxetable}
\normalsize
  

\begin{deluxetable}{lccc}
\tablenum{2}
\tablewidth{0pc}
\tablecaption{Optical/IR Extinction Ratios for R=3.1}
\tablehead{
\colhead{Extinction}            & 
\colhead{Observed}              & 
\colhead{ }                     & 
\colhead{Model Curve}           \\  
\colhead{Ratio}                 & 
\colhead{Value}                 & 
\colhead{Reference\tablenotemark{a}}                  &
\colhead{Value}                 \\ 
\colhead{(1)}  & 
\colhead{(2)}  & 
\colhead{(3)}  & 
\colhead{(4)}  }
\startdata
& & &                                                                    \nl
$A(M)/E(B-V)$   & $\phm{-}0.08-0.12$               & 1,2     & $\phm{-}0.12$   \nl
$A(L)/E(B-V)$   & $\phm{-}0.09-0.20$               & 1,2,3,4 & $\phm{-}0.19$   \nl
$A(K)/E(B-V)$   & $\phm{-}0.33-0.38$               & 2,3,4   & $\phm{-}0.36$   \nl
$A(H)/E(B-V)$   & $\phm{-}0.52-0.55$               & 1,2     & $\phm{-}0.53$   \nl
$A(J)/E(B-V)$   & $\phm{-}0.85-0.91$               & 1,2,3   & $\phm{-}0.86$   \nl
$A(I)/E(B-V)$   & $\phm{-}1.50$                     & 3       & $\phm{-}1.57$   \nl
$A(R)/E(B-V)$   & $\phm{-}2.32$                     & 3       & $\phm{-}2.32$   \nl
$A(V)/E(B-V)$   & $\phm{-}3.10$                     &         & $\phm{-}3.10$   \nl
& & &                                                                    \nl
$E(U-B)/E(B-V)$ & $\phm{-}0.70+0.05\times E(B-V)$ & 5 & $\phm{-}0.69+0.04\times E(B-V)$ \nl
& & &                                                                    \nl
$E(b-y)/E(B-V)$ & $\phm{-}0.74$                    & 6 & $\phm{-}0.74$   \nl
$E(m1)/E(b-y)$  & $-0.32$                          & 6 & $-0.32$         \nl
$E(c1)/E(b-y)$  & $\phm{-}0.20$                    & 6 & $\phm{-}0.17$   \nl
$E(u-b)/E(b-y)$ & $\phm{-}1.5$                     & 6 & $\phm{-}1.54$   \nl
\tablenotetext{a}{References: 1 = Rieke \& Lebofsky 1985;  2 = Whittet 1988; 3 = Schultz \& Wiemer 1975; 4 = Savage \& Mathis 1979; 5 = FitzGerald 1970; 6 = Crawford 1975 }
\enddata
\end{deluxetable}


\begin{deluxetable}{ccc}
\tablenum{3}
\tablewidth{0pc}
\tablecaption{Cubic Spline Anchor Points for R = 3.1 Curve}
\tablehead{
\colhead{Wavelength}                & 
\colhead{$\lambda^{-1}$}            & 
\colhead{\underline{\hspace{1.0em}$A(\lambda)$\hspace{1.0em}}}   \\  
\colhead{(\AA)}                     & 
\colhead{(\micron$^{-1}$)}          & 
\colhead{$E(B-V)$}                  \\ 
\colhead{(1)}                       & 
\colhead{(2)}                       & 
\colhead{(3)}  }
\startdata  
$\infty$      &  0.000   & 0.000   \nl
26500         &  0.377   & 0.265  \nl
12200         &  0.820   & 0.829  \nl
\phn6000      &  1.667   & 2.688  \nl
\phn5470      &  1.828   & 3.055  \nl
\phn4670      &  2.141   & 3.806  \nl
\phn4110      &  2.433   & 4.315  \nl
\phn2700      &  3.704   & 6.265  \nl
\phn2600      &  3.846   & 6.591  \nl
\enddata
\end{deluxetable}


\begin{deluxetable}{ccc}
\tablenum{4}
\tablewidth{0pc}
\tablecaption{R-Dependent Values of Optical Spline Anchor Points}
\tablehead{
\colhead{Wavelength}                & 
\colhead{$\lambda^{-1}$}            & 
\colhead{\underline{\hspace{1.0em}$A(\lambda)$\hspace{1.0em}}}   \\  
\colhead{(\AA)}                     & 
\colhead{(\micron$^{-1}$)}          & 
\colhead{$E(B-V)$}                  \\ 
\colhead{(1)}                       & 
\colhead{(2)}                       & 
\colhead{(3)}  }
\startdata  
6000      &  1.667   & $-0.426 + 1.0044\times R$ \nl
5470      &  1.828   & $-0.050 + 1.0016\times R$ \nl
4670      &  2.141   & $\phm{-}0.701 + 1.0016\times R$  \nl
4110      &  2.433   & $\phm{-}1.208 + 1.0032\times R - 0.00033\times R^2$ \nl
\enddata
\end{deluxetable}
 

\footnotesize
\begin{deluxetable}{lcccccc}
\tablenum{5}
\tablewidth{0pc}
\tablecaption{Comparison of Model IR Color Excesses With Observations}
\tablehead{
\colhead{ }                     & 
\colhead{$E(B-V)$}              & 
\colhead{$E(V-J)$}              & 
\colhead{$E(V-H)$}              & 
\colhead{$E(V-K)$}              & 
\colhead{$E(V-L)$}              &
\colhead{ }                     \\  
\colhead{Sightline}             & 
\colhead{(mag)}                 & 
\colhead{(mag)}                 & 
\colhead{(mag)}                 & 
\colhead{(mag)}                 & 
\colhead{(mag)}                 & 
\colhead{$R$}                   \\ 
\colhead{(1)}                   & 
\colhead{(2)}                   & 
\colhead{(3)}                   & 
\colhead{(4)}                   & 
\colhead{(5)}                   & 
\colhead{(6)}                   & 
\colhead{(7)}                   }
\startdata  
HD 210121 --- observed & 0.38 & 0.64 & 0.71 & 0.73 & 0.78 & \nodata       \nl
HD 210121 --- model    & \nodata & 0.61 & 0.70 & 0.75 & 0.79 & $2.22\pm0.14$ \nl
\nl
HD \phn21483 --- observed & 0.55 & 1.07 &  1.22 &  1.32  & 1.40  & \nodata       \nl
HD \phn21483 --- model    & \nodata & 1.08 &  1.23 &  1.31  & 1.39  & $2.69\pm0.11$ \nl
\nl
HD 167771 --- observed & 0.44 & 0.96 & 1.14 & 1.23 & 1.25 & \nodata       \nl
HD 167771 --- model    & \nodata & 0.98 & 1.13 & 1.20 & 1.27 & $3.10\pm0.16$ \nl
\nl
HD 147889 --- observed & 1.09 & 2.98 & 3.52 & 3.90 & 4.11 & \nodata       \nl
HD 147889 --- model    & \nodata & 3.11 & 3.58 & 3.81 & 4.04 & $3.92\pm0.17$ \nl
\nl
HD \phn37061 --- observed & 0.54 & 1.74 & 2.05 & 2.21 & 2.34 & \nodata       \nl
HD \phn37061 --- model    & \nodata & 1.78 & 2.06 & 2.19 & 2.32 & $4.57\pm0.18$ \nl
\nl
HD \phn38087 --- observed & 0.33 & 1.10 & 1.45 & 1.58 & 1.53 & \nodata         \nl
HD \phn38087 --- model    & \nodata & 1.21 & 1.40 & 1.49 & 1.57 & $5.08\pm0.58$ \nl
\nl
HD \phn36982 --- observed & 0.34 & 1.38 & 1.61 & 1.85 & \nodata & \nodata         \nl
HD \phn36982 --- model    & \nodata & 1.43 & 1.66 & 1.76 & \nodata & $5.83\pm0.57$ \nl
\nl
\tablecomments{Observed color excesses are derived using the intrinsic
colors from Wegner 1994.  Model color excesses are computed via
synthetic photometry of artificially reddened stellar energy
distributions using the IR extinction law derived in the Appendix and
assuming the $E(B-V)$ values from column 2 and the $R$ values from
column 7.  The stated values of $R$ yield the best fits to the observed
IR color excesses and the uncertainties are based on the assumptions
$\sigma_{E(B-V)} = 0.02$ and $\sigma_{E(V-\lambda)} = 0.05$.  Kurucz
ATLAS9 models with appropriate values of $T_{eff}$ were used to
represent the intrinsic stellar energy distributions.} 
\enddata
\end{deluxetable}
\normalsize

%
%

\newpage
\begin{figure}
\plotfiddle{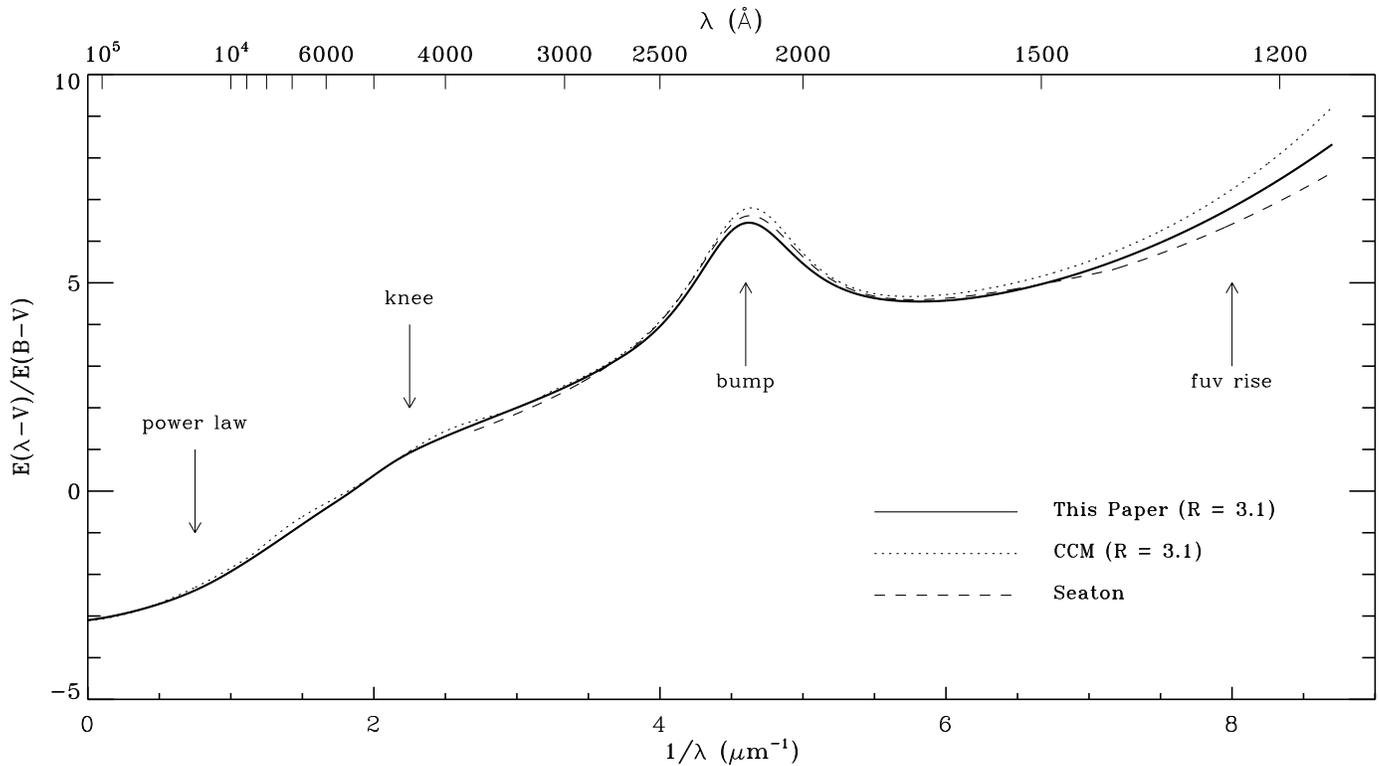}{3.4in}{90}{70}{70}{275}{-65}
\figcaption{Normalized interstellar extinction curves from the far-IR
through the UV.  Several general features of the curves are noted.  The
solid and dotted curves are estimates for the case $R \equiv
A(V)/E(B-V) = 3.1$ derived in the Appendix of this paper and by
Cardelli, Clayton, \& Mathis 1989, respectively.  The dashed
curve shows the average Galactic UV extinction curve from Seaton 1979.}
\end{figure}
 
\begin{figure}
\plotfiddle{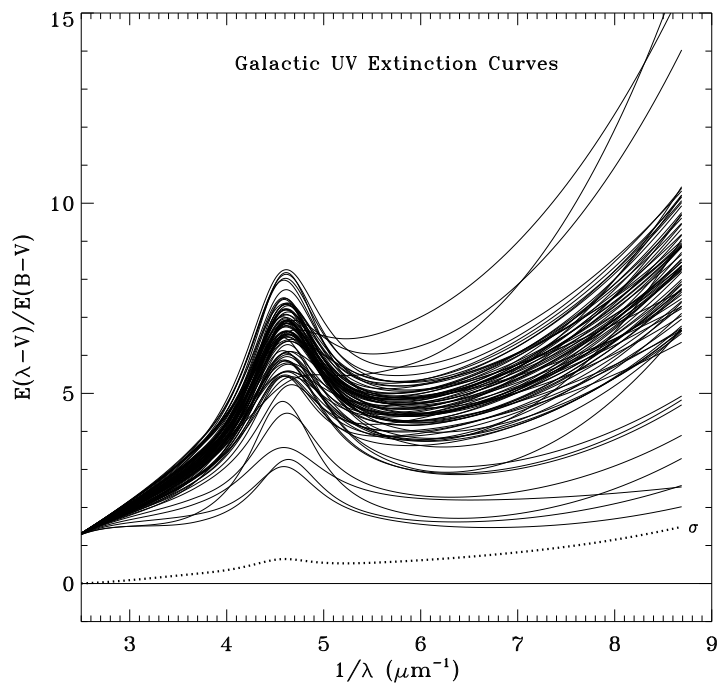}{2.5in}{90}{50}{50}{205}{-23}
\figcaption{Examples of 80 Galactic UV extinction curves derived from {\it
IUE} satellite observations.  Analytical fits to the curves are shown,
based on the work of Fitzpatrick \& Massa 1990.  The curves are taken
from the Fitzpatrick \& Massa catalog, with the addition of the
lines-of-sight toward HD 210121 from Welty \& Fowler 1992 and HD 62542
from Cardelli \& Savage 1988.  This figure demonstrates the enormous
range of properties exhibited by UV extinction in the Milky Way.  The
dotted line, labeled ``$\sigma$,'' shows the standard deviation of the
sample scaled to the value $\sigma(1500 \AA) = 0.74$, as derived from
{\it ANS} satellite data (see \S 3.1).}
\end{figure}
 
\begin{figure}
\plotfiddle{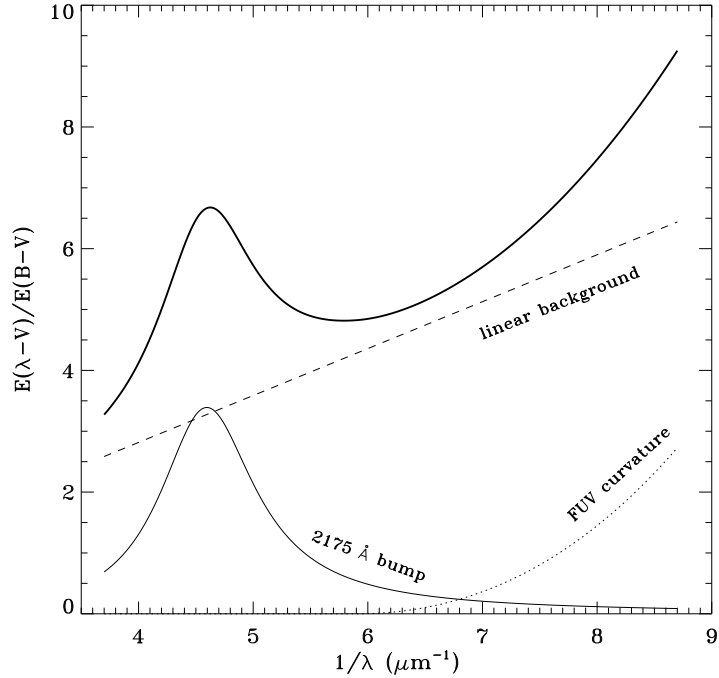}{2.9in}{90}{50}{50}{205}{-23}
\figcaption{Analytical fitting functions for UV extinction curves from
Fitzpatrick \& Massa 1990.  A normalized UV extinction curve (thick
solid curve) can be represented by a combination of three functions:
(1) a linear background component (thin dashed line), (2) a UV bump
component (thin solid curve), and (3) a far-UV curvature component
(thin dotted line).  The linear background is parameterized by two
tightly correlated coefficients (slope and intercept), the bump by
three coefficients (strength, width, and central position), and the
far-UV curvature by a single scale factor.  Given the near invariance
in bump central position and the correlation between the linear
coefficients, most Galactic UV extinction curves can be reproduced to
within the observational uncertainties with only four free parameters.}
\end{figure}
  
\begin{figure}
\plotfiddle{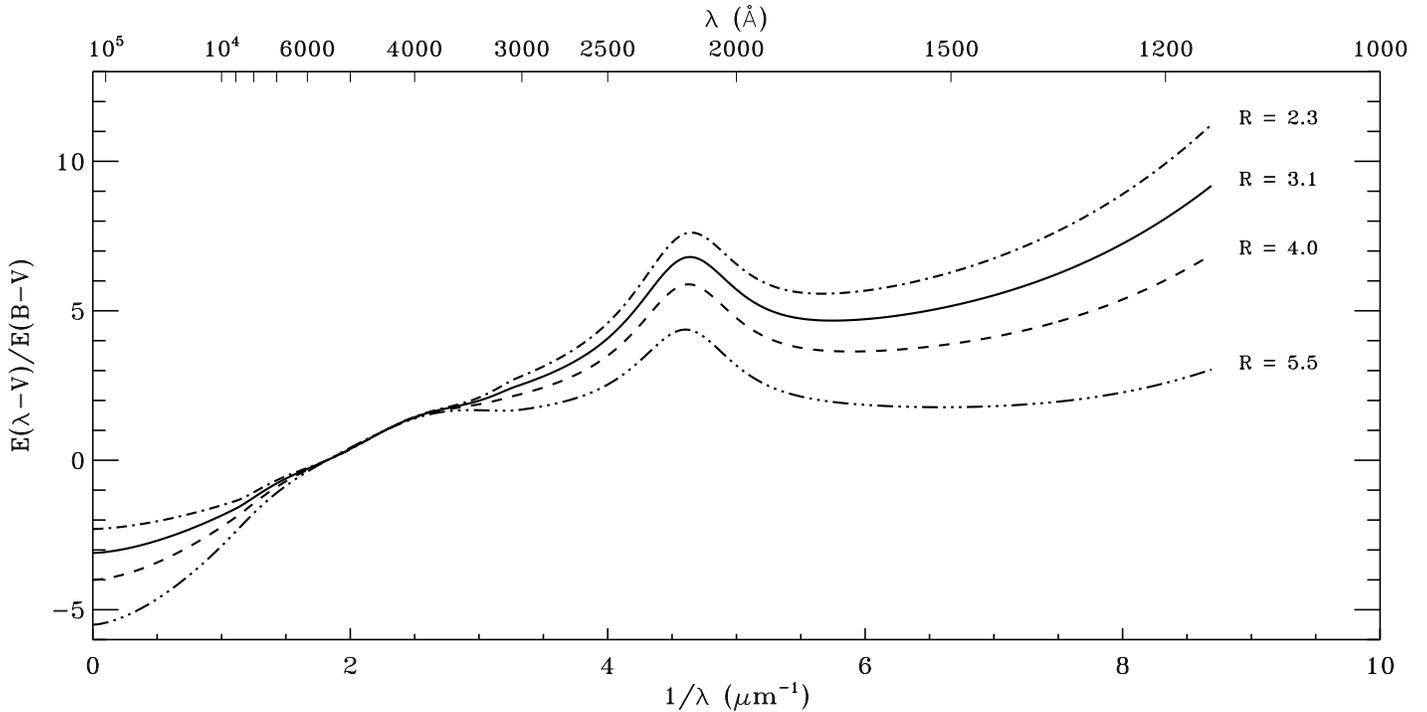}{2.5in}{90}{70}{70}{275}{-79}
\figcaption{Far-IR through UV extinction curves from Cardelli, Clayton,
and Mathis 1989 (CCM).  CCM found that extinction curves can be
expressed approximately as a 1-parameter family that varies linearly
with $R^{-1}$, where $R \equiv A(V)/E(B-V)$ and has a mean value in the
diffuse interstellar medium of 3.1.   Examples of CCM's results are
shown for four representative values of $R$, listed on the righthand side
of the figure next to the corresponding curve.}
\end{figure}
 
\begin{figure}
\plotfiddle{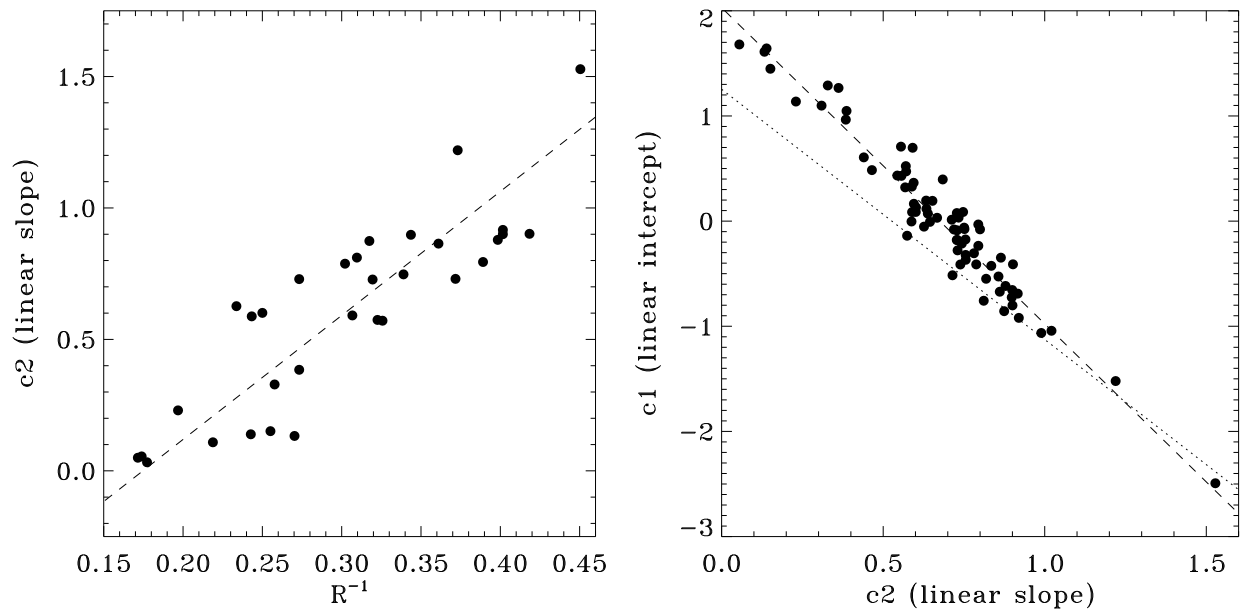}{2.5in}{90}{60}{60}{230}{-50}
\figcaption{{\it Left Panel:}  Slope of the UV linear extinction component
$c2$ plotted against $R^{-1}$ (filled circles) for 31 sightlines from
the Fitzpatrick \& Massa 1990 (FM) sample plus HD 210121 (at $R^{-1} =
0.45$).  The adopted linear relationship between these quantities is
indicated with the dashed line and given by $c2 = -0.824 +
4.717R^{-1}$.   {\it Right Panel:} Intercept of the UV linear
extinction component $c1$ plotted against the linear slope $c2$ (filled
circles) for the full set of $\sim$80 extinction curves from the FM
catalog.  The adopted linear relationship between these parameters is
indicated by the dashed line and given by $c1 =  2.030 - 3.007 c2$. The
relationship between $c1$ and $c2$ implicit in the CCM formula is shown
by the dotted line.}
\end{figure}
 
\begin{figure}
\plotfiddle{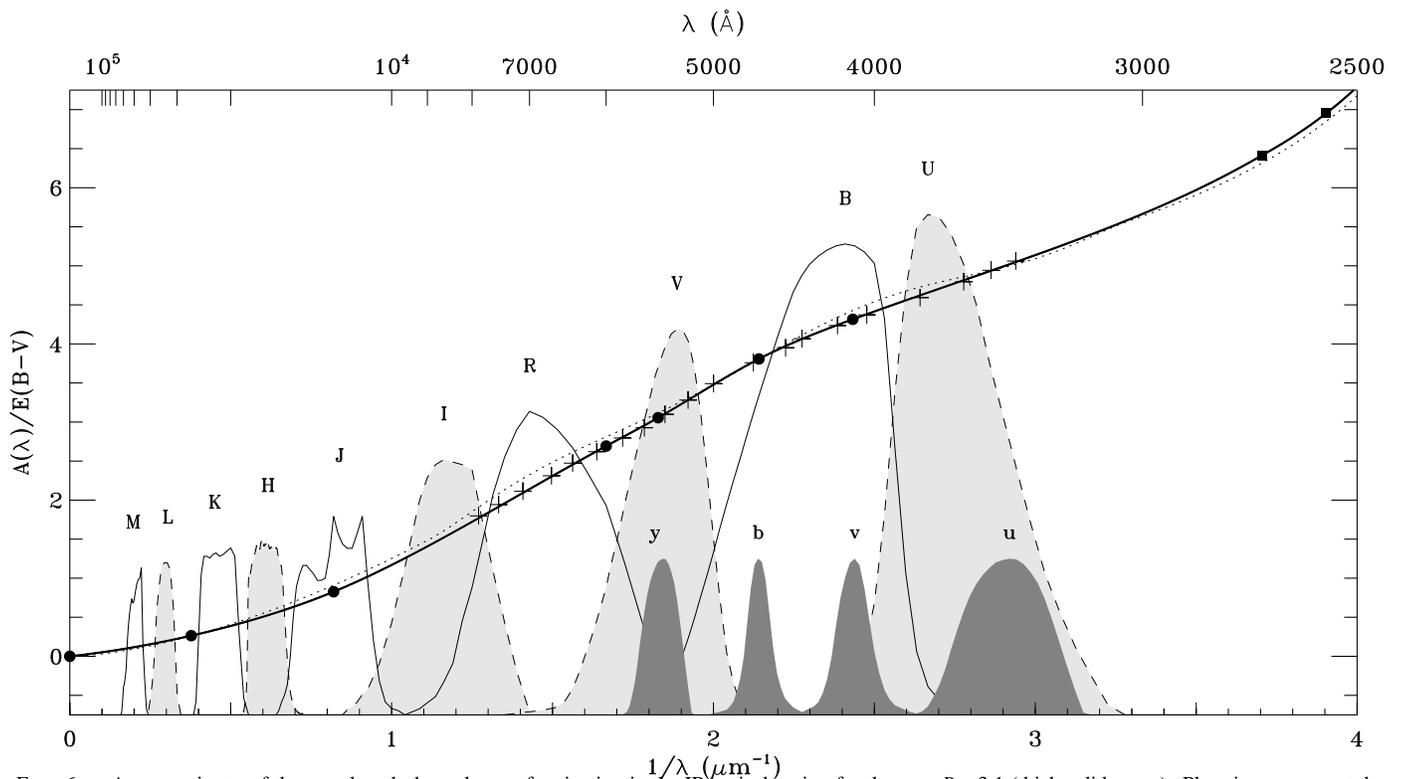}{2.5in}{90}{70}{70}{275}{-79}
\figcaption{A new estimate of the wavelength dependence of extinction in
the IR/optical region for the case $R = 3.1$ (thick solid curve).  Plus
signs represent the extinction data from Bastiaansen 1992, normalized
to $R = 3.1$;  the dotted line shows the CCM curve for $R = 3.1$.  The
arbitrarily scaled profiles of the Johnson {\it UBVRIJHKLM} and
Str\"{o}mgren {\it uvby} filters are shown for comparison.   The new
curve was constrained to reproduce the broad- and intermediate-band
filter-based extinction measurements listed in Table 2 and to fit the
Bastiaansen data.   For $\lambda > 2700$ \AA\/ ($1/\lambda < 3.7
\micron^{-1}$) the curve is constructed as a cubic spline interpolation
between the points marked by the filled symbols (see Table 3). At
wavelengths shortward of 2700 \AA, the curve is computed using the FM
fitting function with the coefficients given in the Appendix.}
\end{figure}
 
\begin{figure}
\plotfiddle{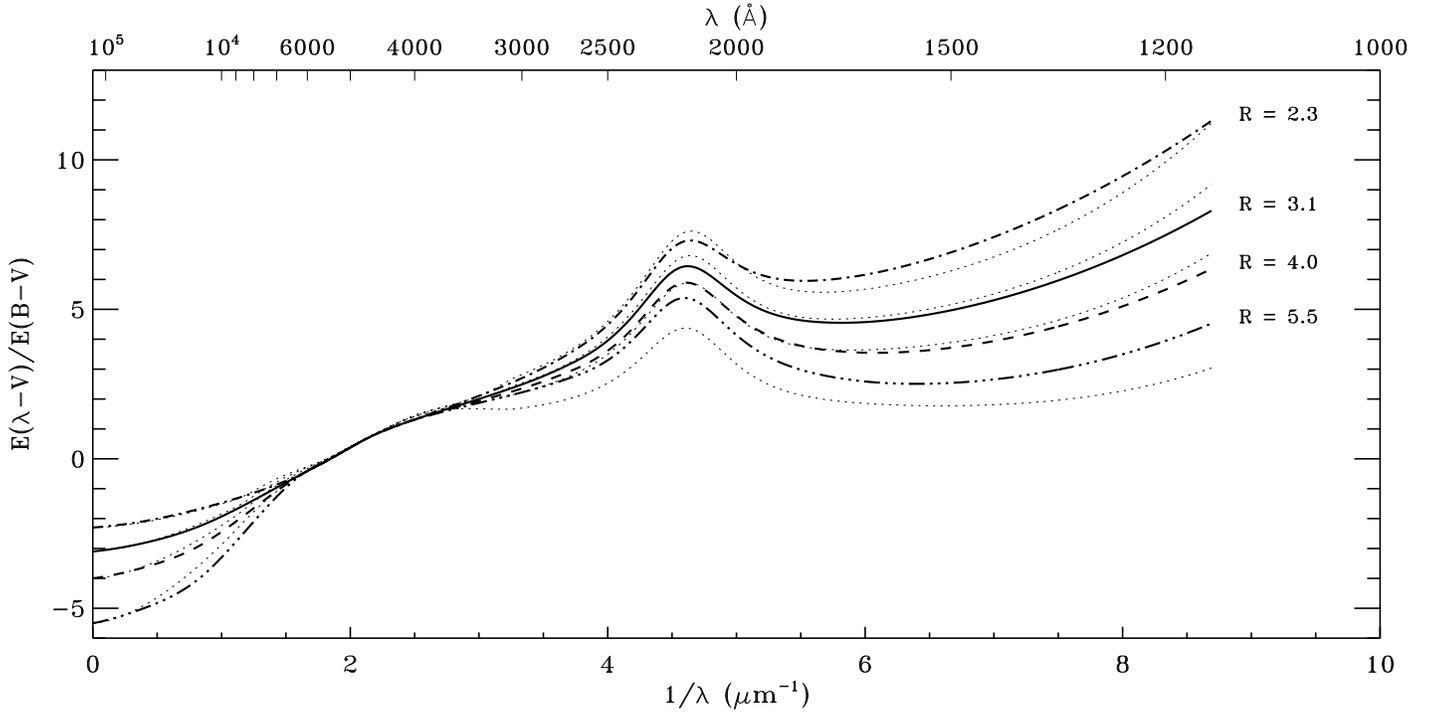}{3.3in}{90}{70}{70}{275}{-79}
\figcaption{Examples of the $R$-dependent far-IR through UV extinction
curves derived in this paper (thick solid and dashed curves).  The
corresponding values of $R$ are listed on the righthand side of the
figure beside the curves.  For comparison, the results of CCM for the
same four values of $R$ are shown.}  
\end{figure}
 
\begin{figure}
\plotfiddle{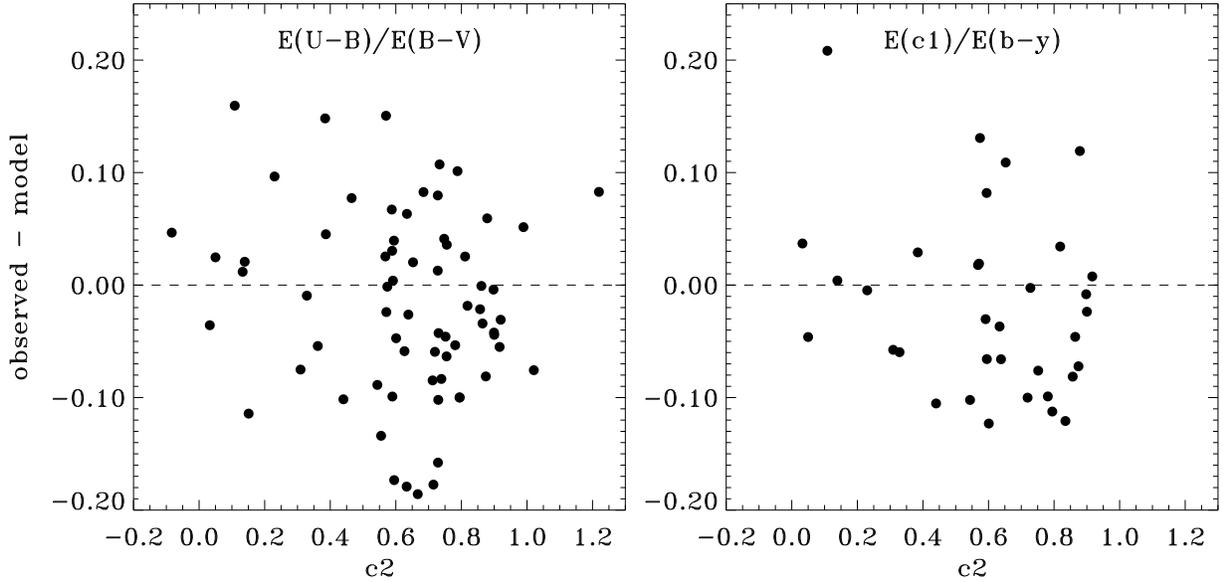}{2.5in}{90}{60}{60}{230}{-50}
\figcaption{{\it Left Panel:} Observed-minus-predicted values of
$E(U-B)/E(B-V)$ plotted against the slope of the UV linear extinction
component $c2$ (filled circles) for sightlnes from the FM sample.
Model values of $E(U-B)/E(B-V)$ vary over the range $\sim$0.6 to
$\sim$0.8 for the observed range in $c2$. {\it Right Panel:}  Observed
minus predicted values of the Str\"{o}mgren extinction ratio
$E(c1)/E(b-y)$ plotted against $c2$ (filled circles) for the FM
sightlines with Str\"{o}mgren data.  Model values of $E(c1)/E(b-y)$
vary over the range $\sim$0.0 to $\sim$0.3 for the observed range in
$c2$.}
\end{figure}
\clearpage

\end{document}